\documentclass{aa} % for a referee version
\usepackage{graphicx}
\usepackage{amsmath}
\usepackage{natbib}
\usepackage{longtable}
\bibpunct{(}{)}{;}{a}{}{,}
\usepackage{txfonts}

\newcommand{\bv}{\mathbf{v}}
\newcommand{\bb}{\mathbf{B}}
\newcommand{\mrho}{\overline{\varrho}}
\newcommand{\mT}{\overline{T}}
\newcommand{\mP}{\overline{P}}
\begin{document}
   \title{Topology and field strength \\
     in spherical, anelastic dynamo simulations}

   \author{M. Schrinner
     \thanks{\emph{Present address:}
       Universit\"at G\"ottingen, Institut f\"ur Astrophysik,
       Friedrich-Hundt-Platz 1, 37077 G\"ottingen, Germany}   
       \and L. Petitdemange\and R. Raynaud \and E. Dormy}

   \institute{MAG (ENS / IPGP), LRA, D\'epartement de Physique,
              \'Ecole Normale Sup\'erieure, 24 rue Lhomond,
              75252 Paris Cedex 05, France\\
   \email{martin@schrinner.eu}}

   \date{Received 4 October 2013 / Accepted 30 December 2013}

% \abstract{}{}{}{}{} 
% 5 {} token are mandatory
 
  \abstract
  % context heading (optional) 
  {Numerical modelling of convection
  driven dynamos in the Boussinesq approximation revealed fundamental
  characteristics of the dynamo-generated magnetic fields and the
  fluid flow.  Because these results were obtained for an
  incompressible fluid of constant density, their validity for gas
  planets and stars remains to be assessed.  A common approach is to
  take some density stratification into account with the so-called
  anelastic approximation.  }
  % aims heading (mandatory)
   {The validity of previous results obtained in the Boussinesq
     approximation is tested for anelastic models. We point out and
     explain specific differences between both types of models, in
     particular with respect to the field geometry and the field
     strength, but we also compare scaling laws for the velocity
     amplitude, the magnetic dissipation time, and the convective heat
     flux. }
  % methods heading (mandatory)
   {Our investigation is based on a systematic parameter study of spherical 
     dynamo models in the anelastic approximation. We make use of a recently
     developed numerical solver and provide results for the test cases of the 
     anelastic dynamo benchmark.
   }
  % results heading (mandatory)
   {The dichotomy of dipolar and multipolar dynamos identified in Boussinesq
     simulations is also present in our sample of anelastic
     models. Dipolar models require that the typical length scale of
     convection is an order of magnitude larger than the Rossby radius.
     However, the distinction between both classes of models is
     somewhat less explicit than in previous studies. This is mainly
     {due to} two 
     reasons: we found a number of models with a considerable equatorial 
     dipole contribution and an intermediate overall dipole field strength. 
     Furthermore, a large density stratification may hamper the generation of
     dipole dominated magnetic fields. {Previously proposed scaling} laws, 
     such as those for the field strength, are similarly applicable to anelastic models. 
     {It is not clear, however, if this consistency 
     necessarily implies similar dynamo processes in both settings.}
    }     
   {}

   \maketitle
%
%________________________________________________________________
\section{Introduction}
Magnetic fields of low-mass stars and planets are maintained by
currents resulting from the motion of a conducting fluid (or gas) in
their interiors. Because the magnetic field acts back on the flow via
the Lorentz force, the hydrodynamic dynamo problem is intrinsically
nonlinear. Moreover, a sufficiently complex flow and magnetic field
geometry has to be assumed in order to enable dynamo action. Further
complications result from tiny diffusivities, such as small kinematic
viscosities, which introduce small dynamic length scales compared to
stellar or planetary radii. Thus, self-consistent simulations of
natural dynamos are not only three-dimensional, but a vast range of
spatial and temporal scales has to be resolved. These difficulties
{prevented} a direct numerical treatment of the dynamo problem for {a
  significant} time.  Only for the last 20 years increasing computer
power made global, direct numerical simulations feasible, in
particular for the geodynamo problem
\citep[e.g.][]{glatzroberts95,kageyama97,kuang97,christensen98,sarson98,
  katayama99,buffet00,dormy2000}; for an early cylindrical annulus
model of the geodynamo see \cite{busse75}. In these simulations, an
incompressible conducting fluid was considered, and the Boussinesq
approximation was applied.  Intensive and systematic parameter studies
revealed fundamental properties of these models related to their field
and flow topologies, their field strength, their velocity amplitudes,
their advective heat transport, or their time dependence
\citep[e.g.][]{COG99,grote2000,kutzner02,busse06,christensen06,
  sreenivasan06,busse09,hori2010,landeau2010,schrinner12,yadav12}.
The simplifying assumption of constant density in Boussinesq models,
however, is probably not justified for gas planets or stars, {in
  which} the density typically varies over many scale heights. An
alternative approach, which takes compressibility into account, is the
so-called anelastic approximation \citep{ogura62,gough69,gilman81}. In
anelastic models, the density varies with radius, but {its time
  derivative is neglected in the continuity equation} and the mass
flux is solenoidal
\citep[e.g.][]{G84,braginsky95,lantz99,miesch2000,brun04,
  browning04}. Consequently, fast travelling sound waves are filtered
out and, compared to fully compressible models, larger time steps in
the discretisation scheme may be reached. In this article, we carry
out a systematic parameter study of global dynamo simulations in the
anelastic approximation guided by well known results {of} Boussinesq
models. In this way, we intend to point out specific differences
between anelastic and Boussinesq models and assess the validity of
previous findings obtained in the Boussinesq approximation. A similar
approach was followed by \cite{gastine12} and \cite{yadav13}. We
compare our results with their findings and discuss some differences
we obtained for our varied sample of models.  The conditions for the
generation of large-scale, dipolar fields
(Section~\ref{sec_field_top}) and the test of the flux-based scaling
law for the magnetic field strength (Section~\ref{sec_scaling_laws}),
originally proposed by \cite{christensen06} for Boussinesq models, are
revisited. We argue that the typical length scale of convection
relative to the Rossby radius is of crucial importance for the
resulting field topology \citep[see also][]{schrinner12} and show that
larger magnetic Prandtl numbers are required to obtain dipolar
solutions with increasing density contrast. Furthermore, the
flux-based scaling laws derived for Boussinesq models seem to hold in
the anelastic approximation as well. However, because of their general
validity, the flux-based scaling laws might not be appropriate to
distinguish between different conditions for magnetic field
generation.

The paper is organised as follows: Section~\ref{sec_dyn_calc} introduces the
anelastic models considered here and the recently developed numerical solver;
results for a numerical benchmark are given in Appendix \ref{benchmark}.
In Section~\ref{sec_field_top}, we present new evidence for the existence of a
class of dynamos dominated by an axial dipole and a class of models with a 
more variable magnetic field geometry. Various scaling laws originally derived
for Boussinesq models are tested and discussed 
in Section~\ref{sec_scaling_laws}, and we give some conclusions in 
Section~\ref{sec_conclusions}.

\section{Dynamo calculations}
\label{sec_dyn_calc}
\subsection{The anelastic approximation}
Convection of a gas or a compressible fluid in the interior of planets and 
stars takes place on a vast range of spatial and temporal scales. Sound waves 
excited in convection zones, for example, have very short oscillation periods 
compared to the turnover time of convection or the magnetic diffusion time 
relevant for the generation of magnetic fields. Thus, extremely small 
timesteps would be required to resolve these waves in numerical dynamo
models. To avoid this problem, simplifications of the governing equations are 
often applied. The anelastic approximation used in this study 
advantageously filters out sound waves \citep{ogura62,gough69,gilman81}; it 
is motivated by the idea 
that the superadiabatic temperature gradient driving convection in planetary or 
stellar convection zones is tiny. The thermodynamic variables are then 
decomposed into the sum of (close to adiabatic) reference values, denoted here 
by an overbar, and perturbations, denoted by a prime,
\begin{equation}
\varrho=\mrho+\varrho',\quad T=\mT+T',\quad P=\mP+P' \, .
\end{equation}
Subsequently, the anelastic equations result from the `thermodynamic
linearization' around the reference state. It should be stressed that
a number of different formulations of the anelastic problem can be
found in the literature
\citep[e.g.][]{G84,braginsky95,lantz99,miesch2000,brun04,rogers05,
  jones09,alboussiere13}. We follow here the approach introduced by
\cite{lantz99} and \cite{braginsky95}, also known as
LBR-approximation. They noticed that the only relevant thermodynamic
variable in the equation of motion is the entropy, if the reference
state is assumed to be close to adiabatic. Further advantages of the
LBR-equations over others are that they give a mathematically
consistent, asymptotic limit of the full, general equations
\citep{jones11} and guarantee the conservation of energy
\citep{brown12}. Moreover, the LBR-equations were used to formulate
anelastic dynamo benchmarks \citep{jones11}. The presentation of the
equations given here follows the benchmark paper and \cite{jones_lin}.

\subsection{Basic assumptions}
We consider a perfect, electrically conducting gas in a rotating
spherical shell with an inner boundary at \(r=r_i\) and an outer
boundary at \(r=r_o\).  The aspect ratio of the shell is then defined
by \(\chi=r_i/r_o\). Convection in our simulations is driven by an
imposed entropy difference, \(\Delta s\), between the inner and the
outer boundary. As discussed above, \(\Delta s\) is assumed to be
small. This implies small convective velocities compared to the speed
of sound. For consistency, we also require that the Alfv\'en velocity
of the magnetic field is small. Moreover, the kinematic
viscosity~\(\nu\), the thermal diffusivity~\(\kappa\) and the magnetic
diffusivity~\(\eta \) are constants throughout in this
paper. Following \cite{jones11} we represent the heat flux in our
models in terms of the entropy gradient instead of the temperature
gradient. This assumption relies on wide-spread ideas about
{turbulent mixing} \citep{braginsky95} but does not follow from
first principles.  Applying this simplification allows us to consider
the entropy as the only relevant thermodynamic variable in the
formulation of the anelastic problem.

\subsection{The reference state}
The reference state of our models is a solution of the hydrostatic equations 
for an adiabatic atmosphere. Moreover, the centrifugal acceleration is 
neglected and we assume that gravity varies 
radially, \(\mathbf{g}=-GM\mathbf{\hat{r}}/r^2\) with G being the 
gravitational constant and  \(M\) the central mass of the star or the planet. 
This admits a polytropic solution for the reference atmosphere,
\begin{equation}
\mP=P_c\,w^{n+1},\quad\mrho=\varrho_c\,w^n,\quad \mT=T_c\,w,\quad 
w=c_0+\frac{c_1 d}{r}, 
\label{ref_state}
\end{equation}
with the polytropic index~\(n\) and \(d=r_o-r_i\). We note that \(n\)
defines the value of the adiabatic exponent \(\gamma\), or the ratio
of specific heats~\(c_p/c_v\), via \(\gamma=(n+1)/n\). The values
~\(P_c\), \(\varrho_c\), and \(T_c\) are taken midway between the
inner and the outer boundary and serve as units for the
reference-state variables. Moreover, the constants~\(c_0\) and \(c_1\)
in (\ref{ref_state}) are defined as
\begin{equation}
c_0=\frac{2w_0-\chi-1}{1-\chi},\quad c_1=\frac{(1+\chi)(1-w_o)}{(1-\chi)^2},
\label{c0_c1}
\end{equation}
with 
\begin{equation}
w_0=\frac{\chi+1}{\chi\exp(N_\varrho/n)+1},\quad w_i=\frac{1+\chi-w_o}{\chi},
\label{w1_w0}
\end{equation}
and \(N_\varrho=\ln{(\varrho_i/\varrho_o)}\), where \(\varrho_i\) 
and \(\varrho_o\) denote the reference state density at the inner and outer 
boundary, respectively. We emphasise again that convection in our models is 
not driven by the reference state, or by the choice of a particular
polytropic index~\(n\), but by an imposed entropy difference~\(\Delta s\) 
between the boundaries.

\subsection{The non-dimensional equations}
The use of non-dimensional equations minimizes the number of free
parameters and is a prerequisite for a systematic parameter study. We
choose the shell width \(d\) as the fundamental length scale of our
models, time is measured in units of \(d^2/\eta\), and \(\Delta s\)
is the unit of entropy. The magnetic field is then measured in units
of \(\sqrt{\Omega\varrho_c\mu\eta}\) where \(\Omega\) is the rotation
rate and \(\mu\) the magnetic permeability.  Finally, our dynamo
models are solutions for the velocity~\(\bv\), the magnetic
field~\(\bb\), and the entropy~\(s\) of the following, non-dimensional
equations,
\begin{eqnarray}
\frac{\partial\bv}{\partial t}+\bv\cdot\nabla\bv & = & 
Pm\,\big[-\frac{1}{E}\nabla\frac{P'}{w^n}
+\frac{Pm}{Pr}Ra\frac{s}{r^2}\mathbf{\hat{r}}-\frac{2}{E}\,\mathbf{\hat{z}}
\times\bv\nonumber\label{mhd1}\\
&&+\mathbf{F}_\nu+\frac{1}{E\,w^n}(\nabla\times\bb)\times\bb\big]\, ,\\
\frac{\partial\bb}{\partial t} & = &\nabla\times(\bv\times\bb)+\nabla^2\bb
\, ,\\
\label{mhd2}
\frac{\partial s}{\partial t}+\bv\cdot\nabla s&=&
w^{-n-1}\frac{Pm}{Pr}\nabla\cdot\left(w^{n+1}\,\nabla s\right)\nonumber\\
&&+\frac{Di}{w}\left[E^{-1}w^{-n}(\nabla\times \bb)^2+Q_\nu\right]\, ,\label{mhd3}\\
\nabla\cdot \left(w^n\bv \right)& = & 0\, ,\label{mhd4}\\
\nabla\cdot\bb & = & 0\, .\label{mhd5}
\end{eqnarray}
In (\ref{mhd1}) we used the viscous force 
\(\mathbf{F}_\nu=w^{-n}\nabla\mathbf{S}\) with the rate of strain tensor
\begin{equation}
  S_{ij}=2w^n\left(e_{ij}-\frac{1}{3}\delta_{ij}\nabla\cdot \bv\right),\quad
  e_{ij}=\frac{1}{2}\left(\frac{\partial v_i}{\partial x_j}+\frac{\partial
    v_j}{\partial x_i}\right) \, .
\end{equation}

Moreover, the dissipation parameter~\(Di\) and the viscous
heating~\(Q_\nu\) in (\ref{mhd3}) are given by
\begin{equation}
Di=\frac{c_1Pr}{Pm Ra} \, ,
\end{equation} 
and
\begin{equation}
Q_\nu=2\left[e_{ij}e_{ij}-\frac{1}{3}(\nabla\cdot\bv)^2\right] \, .
\end{equation}
The system of equations (\ref{mhd1})--(\ref{mhd5}) is governed by a
number of dimensionless parameters. These are the Rayleigh
number~\(Ra\), the Ekman number~\(E\), the Prandtl number~\(Pr\), and
the magnetic Prandtl number~\(Pm\). Together with the aspect
ratio~\(\chi\), the polytropic index~\(n\), and the number of density
scale heights~\(N_\varrho\) defining the reference state, our models
are therefore fully determined by seven dimensionless parameters:
\begin{eqnarray}
Ra&=&\frac{GMd\Delta s}{\nu\kappa c_p},\quad Pr=\frac{\nu}{\kappa},\quad
Pm=\frac{\nu}{\eta},\quad E=\frac{\nu}{\Omega d^2} \, ,\nonumber\\
N_\varrho&=&\ln\left(\frac{\varrho_i}{\varrho_o}\right),\quad n,\quad 
\chi=\frac{r_i}{r_o} \, .
\label{parameter}
\end{eqnarray}

{Following \cite{jones_lin}, we also considered a linearized form of 
equations (\ref{mhd1}) and (\ref{mhd3}) to calculate some critical Rayleigh 
numbers for the onset of convection. These are listed in Table \ref{tab:rac}.}

\subsection{Boundary conditions}
The mechanical boundary conditions are impenetrable and stress free on both 
boundaries,
\begin{equation}
v_r=\frac{\partial}{\partial r}
\left(\frac{v_\theta}{r}\right)=\frac{\partial}{\partial
  r}\left(\frac{v_\phi}{r}\right)=0\qquad\text{on}\quad
r=r_i\quad\text{and}\quad r=r_o \, . 
\label{mech_bound}
\end{equation}
Furthermore, the magnetic field matches a potential field outside the fluid
shell. The choice of these boundary conditions requires that the total angular 
momentum is conserved \citep{jones11}. Finally, the entropy is fixed on the 
inner and the outer boundary with
\begin{equation}
s=\Delta s\qquad\text{on}\quad r=r_i,\quad\text{and}\quad s=0\qquad\text{on}
\quad r=r_o \, .
\label{ent_bound}
\end{equation} 
\subsection{Output parameters}
We use a number of non-dimensional output parameters to characterize our 
numerical dynamo models. These are mostly based on the kinetic and magnetic 
energy densities,
\begin{equation}
E_\mathrm{k}=\frac{1}{2\, V}\int_V\,w^n\bv^2\,\mathrm{d}V \, \,\,
%\label{ekin}
\mbox{and}
\qquad
E_\mathrm{m}=\frac{1}{2\, V}\frac{Pm}{E}\int_V\,\bb^2\,\mathrm{d}V
\, ,
\label{emag}
\end{equation}
where the integrals are taken over the volume of the fluid shell~\(V\).
A non-dimensional measure for the velocity amplitude is then the magnetic 
Reynolds number, \(Rm=\sqrt{2E_\mathrm{k}}\), or the Rossby number, 
\(Ro=\sqrt{2E_\mathrm{k}} E/Pm\). To distinguish models with different field 
geometries it turned out to be useful to introduce also a local Rossby number,
\(Ro_\ell=Ro_c\,\ell_c/\pi\). Here, \(\ell_c\) stands for
the mean harmonic degree of the velocity component \(\bv_c\) from which 
the mean zonal flow has been subtracted \citep{schrinner12},
\begin{equation}
\ell_c=\sum_\ell
\ell\frac{<w^n(\bv_c)_\ell\cdot(\bv_c)_\ell>}{<w^n\bv_c\cdot\bv_c>}
\, .
\label{local_rossby}
\end{equation} 
The brackets in (\ref{local_rossby})  denote an average over time and radii. 
Also, \(Ro_c\) is adapted consistently and stands for the Rossby number based on
the kinetic energy density without the contribution from the mean zonal flow.
{
The definition of \(Ro_\ell\) given here is different from
\cite{christensen06}, as it}
is not based on the total velocity and tries to avoid any dependence on 
the mean zonal flow. 
  
The amplitude of the average magnetic field in our simulations is
measured in terms of the Lorentz number, \(Lo=\sqrt{2E_\mathrm{m}}E/Pm\), 
which was previously used to derive a power law for the field strength in 
Boussinesq simulations \citep{christensen06}. The topology of the field is 
characterized by the relative dipole field strength, \(f_\mathrm{dip}\), 
defined as the time-average ratio on the outer shell boundary of the dipole 
field strength to the total field strength.

The total amount of heat transported in and out of the fluid shell
relative to the conductive heat flux is quantified by the Nusselt number,
\begin{eqnarray}
Nu_\mathrm{bot} & = & -\frac{(\exp{(N_\varrho)}-1)w_ir_i^2}{4\pi
  nc_1}\int_{S_i}\frac{\partial s}{\partial r}\sin\theta\,\mathrm{d}\theta
\mathrm{d}\phi \, ,\label{nu_bot}\\
Nu_\mathrm{top} & = & -\frac{(1-\exp{(-N_\varrho)})w_or_o^2}{4\pi
  nc_1}\int_{S_o}\frac{\partial s}{\partial r}\sin\theta\,\mathrm{d}\theta
\mathrm{d}\phi \, . \label{nu_top}
\end{eqnarray}
The integrals are taken here over the spherical surface at radius \(r_i\) and 
radius \(r_o\), respectively. {For a steady equilibrium state, 
\(Nu_\mathrm{bot}\) and \(Nu_\mathrm{top}\) are identical if time averaged.}
For later use, we also define a Nusselt number based on the advective heat flux
alone, 
\begin{equation}
Nu^\star=(Nu_\mathrm{bot}-1)\frac{E}{Pr} \, ,
\label{def_nu}
\end{equation}
and accordingly a {quantity usually referred as the} flux based Rayleigh number, 
\begin{equation}
Ra_Q=(Nu_\mathrm{bot}-1)\,\frac{Ra\, E^3}{r_o^2\, Pr^2} \, . 
\label{def_raq}
\end{equation}
The energy balance plays a crucial role in {the classical derivation of} 
scaling laws for the saturation level of the magnetic field. In particular, 
the fraction of ohmic to total dissipation, \(f_\mathrm{ohm}=D/P\), 
is {introduced} because it determines the available power used for the 
magnetic field generation. In an equilibrium state, the total dissipation 
equals the power released by buoyancy, 
\begin{equation}
P=\frac{Ra\,E^3}{Pr\,Pm}\int_V\,w^n\,v_r\,s\,\mathrm{d}V \, , 
\label{power}
\end{equation}
and the ohmic dissipation is given by
\begin{equation}
D=\left(\frac{E}{Pm}\right)^2\int_V\,(\nabla\times\bb)^2\,\mathrm{d}V
\, .
\label{dissipation}
\end{equation}
In  (\ref{power}) and (\ref{dissipation}), we scaled \(P\) and \(D\) by
\(\varrho_c\Omega^3d^5\). 

\subsection{Equation for a tracer field}
For some of our models, we calculated the evolution of a magnetic
tracer field simultaneously with (\ref{mhd1})--(\ref{mhd5}),
\begin{equation}
\frac{\partial\bb_\mathrm{Tr}}{\partial
  t}=\nabla\times(\bv\times\bb_\mathrm{Tr})+\nabla^2\bb_\mathrm{Tr} \, .
\label{tracer}
\end{equation}
In the above equation, \(\bb_\mathrm{Tr}\) is a passive vector field which was
advanced at each time step kinematically with the quenched velocity field, 
but \(\bb_\mathrm{Tr}\) did not contribute to the 
Lorentz force \citep{cattaneo09b,tilgner08,schrinner10a}. 
\cite{schrinner10a} found that \(\bb_\mathrm{Tr}\) grows exponentially for
multipolar dynamos but stays stable for models dominated by a dipole field.
The stability of \(\bb_\mathrm{Tr}\) will also serve in this study to
distinguish between different classes of dynamos \citep{schrinner12}.

\subsection{Numerical implementation}
The numerical solver used to compute solutions of
equations~(\ref{mhd1})--(\ref{mhd5}) is a recently developed,
anelastic version of PaRoDy (\cite{dormy98} and further
developments). The code uses a poloidal-toroidal expansion and a
pseudo-spectral spherical harmonic expansion. The numerical method is
similar in these aspects to the one originally introduced in
\cite{G84}. The radial discretisation, however, is based on finite
differences on a stretched grid (allowing for a parallelization by a
radial domain decomposition). Moreover, the pressure term has been
eliminated by taking twice the curl of the momentum equation.  The
anelastic benchmark results obtained with PaRoDy are presented in
Appendix \ref{benchmark}.

\section{Field topology}
\label{sec_field_top}
\subsection{Dipolar and multipolar dynamos}
Parameter studies for Boussinesq simulations revealed two distinct
classes of dynamo models. They can be distinguished by their field
geometry and are therefore referred to as `dipolar' and `multipolar'
models \citep{kutzner02,christensen06}. The spatial variability of
multipolar dynamos is a direct consequence of dynamo action in a
turbulent environment and has to be expected. The class of dipolar
dynamos, however, is more peculiar. \cite{schrinner11c} showed that
these models are single-mode dynamos, that is, except for the
fundamental mode, all more structured magnetic eigenmodes are highly
damped. The single-mode property leads to further characteristic
differences between both classes of dynamos, apart from their
different field geometries. Whereas the dipole axis is stable for
models with a dominant axial dipole field, multipolar models show
frequent polarity reversals \citep{kutzner02} or oscillations
\citep{goudard08,schrinner12}. A third fundamental difference between
dipolar and multipolar models is related to their saturation
mechanism. If a magnetic tracer field is advanced kinematically with
the self-consistent, quenched velocity field stemming from the full
dynamo simulation, the tracer field grows exponentially for multipolar
but not for dipolar models. Dipolar dynamos are ``kinematically stable''
and in this numerical experiment, the tracer field becomes aligned
with the actual, self-consistent magnetic field after some initial
transition period \citep{schrinner10a}.  Finally, dipolar and
multipolar dynamos follow slightly different scaling laws for the
magnetic field \citep{christensen10,schrinner12,yadav12}. This aspect
will be further discussed in Section~\ref{sec:mag_field}.

\cite{christensen06} proposed a criterion based on a local Rossby number to
separate dipolar from multipolar dynamos. We adopt this criterion in a
slightly altered form \citep{schrinner12}. It says that dipolar dynamos may 
be found if the typical length scale of convection, \(\ell\), is at least 
an order of magnitude larger than the Rossby radius, or,
\(Ro_\ell={\rm v}/(\Omega\ell)<0.12\) (in which \({\rm v}\) is a typical rms 
velocity). Our criterion is different from \cite{christensen06}, and entirely 
based on convection and not influenced by the mean zonal flow. This helped to 
generalize the Rossby number rule to models with different aspect ratios and 
mechanical boundary conditions \citep{schrinner12}. Moreover, our 
reinterpretation assumes that the magnetic field is generated only by 
convection and therefore explains why the Rossby number criterion is not 
applicable to models for which differential rotation plays an essential role. 
\begin{figure}
  \centering
  \includegraphics[scale=0.6]{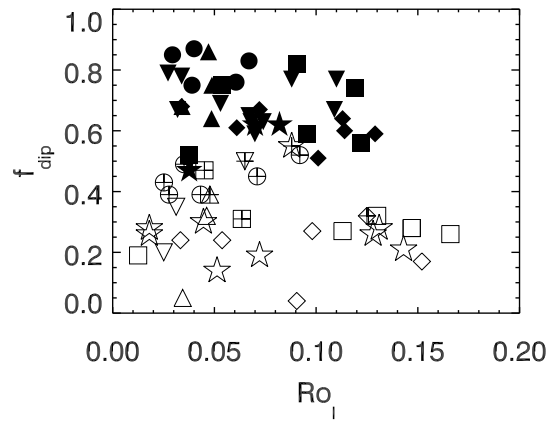}
  \caption{The relative dipole field-strength versus the local Rossby number
  for our sample of models. Filled symbols stand for dipolar, open symbols
  for multipolar dynamos. The symbol shape indicates the number of density
  scale heights: \(N_\varrho=0.5\): circle; \(N_\varrho=1\): upward triangle; 
  \(N_\varrho=1.5\): downward triangle; \(N_\varrho=2\): diamond;
  \(N_\varrho=2.5\): square; \(N_\varrho=3,3.5,4\): star. A cross inscribed in 
  some open symbols means that the field of these models exhibits a strong 
  equatorial dipole component.
  }
  \label{fig_rol_dip}
\end{figure} 

Figure~\ref{fig_rol_dip} shows the relative dipole field strength versus 
the local Rossby number for all anelastic models considered here. 
\cite{gastine12} presented a similar plot but with \(f_\mathrm{dip}\) based on 
the magnetic energy density instead of the field strength. This leads to
considerably lower values of \(f_\mathrm{dip}\) for multipolar dynamos. 
As for Boussinesq simulations, only multipolar 
models are found for \(Ro_\ell>0.12\) \citep{christensen06}, and the 
multipolar branch extends into the dipolar regime in the form of a bistable 
region where both solutions are possible depending on the initial 
conditions \citep{schrinner12}. However, in 
contrast to comparable parameter studies of Boussinesq models
\citep{christensen06,schrinner12}, dipolar and multipolar dynamos are hardly
distinguishable from each other in terms of their relative dipole field 
strength. Contrary to previous results, models with an 
intermediate dipolarity (\(f_\mathrm{dip}\approx 0.5\)) lead to a fairly smooth 
transition of \(f_\mathrm{dip}\) in Fig.~\ref{fig_rol_dip}. These are in 
particular those models with a high equatorial dipole contribution denoted by 
a cross inscribed in the plotting symbol. Because the dipole field strength 
alone is not conclusive to classify our models in Fig.~\ref{fig_rol_dip}, 
their time-dependence, their kinematic stability, and their scaling 
behaviour (see Sect. \ref{sec:mag_field}) were additionally considered to 
assign them to one of both classes. 

\begin{figure}
  \centering
  \includegraphics[scale=0.8]{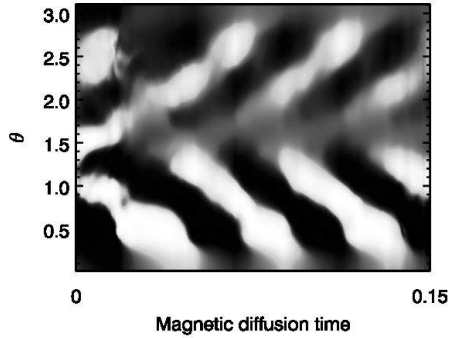}
  \caption{Contour plot of the azimuthally averaged radial magnetic field 
    of model3m versus time and colatitude. The contour plot was normalised 
    by the maximum absolute value at each time step. The grey-scale coding 
    ranges from -1, white, to +1, black.
  }
  \label{fig_butterfly}
\end{figure}

As in the case of Boussinesq simulations, only multipolar models were
found to exhibit polarity reversals or oscillatory dynamo solutions. An 
example of a coherent dynamo wave for model3m (\(N_\varrho=3\)) is given 
in Fig.~\ref{fig_butterfly}. The period of these oscillatory dynamo modes and 
the poleward propagation direction of the resulting wave can be surprisingly 
well explained by Parker's plane layer formalism 
\citep{parker55,busse06,goudard08,schrinner11a,gastine12}. However, the recent 
claim that dynamo waves could migrate towards the equator 
if there is a considerable density stratification \citep{kaepylae13} was not 
confirmed by our simulations. 

\begin{figure}
  \centering
  \includegraphics[scale=0.6]{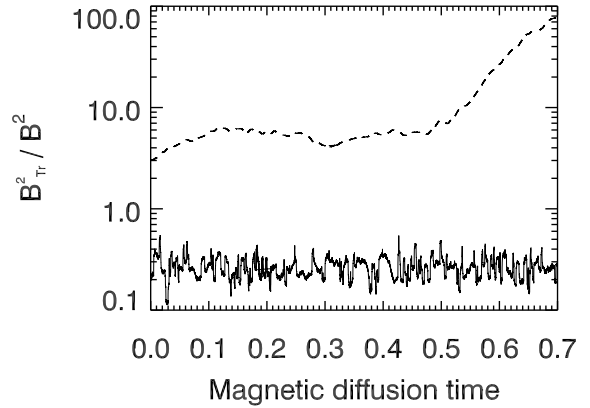}
  \caption{Evolution of the energy of the tracer field normalised by the 
    energy of the actual magnetic field for model2m (dashed line) and 
    model54d (solid line).}
  \label{fig_kin_stability}
\end{figure}

Moreover, we tested 13 arbitrarily chosen models (see the caption of
Table \ref{tab:2}) for kinematic stability and found the dipolar
models to be kinematically stable, whereas all multipolar models
considered exhibited at least periods of
instability. Figure~\ref{fig_kin_stability} shows as an example the
evolution of the kinematically advanced tracer field for model2m and
model54d. For the first, the tracer field grows exponentially but it
stays stable for the latter although it has been permanently perturbed
during the simulation.

\begin{figure}
  \centering
  \includegraphics[scale=0.6]{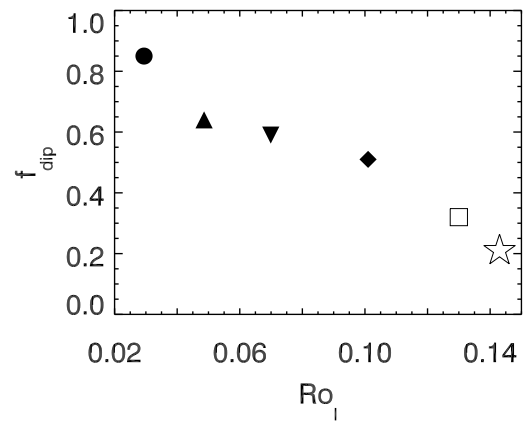}
  \caption{Relative dipole field strength versus \(Ro_\ell\) for a sequence
    of models with \(E=3\times 10^{-4}\), \(Ra=4Ra_c\), \(Pm=3\), and
    \(Pr=1\). The meaning of the symbols is as defined in the caption of 
  Fig.~\ref{fig_rol_dip}.}
  \label{fig_fdip_nrho}
\end{figure} 

A transition from the dipolar to the multipolar regime can be triggered by a
decrease in the rotation rate or the {dynamical} length scale
{(possibly associated with a change in the aspect ratio)}, or an increase in
the velocity amplitude. These three quantities influence the local Rossby
number directly. In Fig. \ref{fig_fdip_nrho}, we show that a transition towards
the multipolar regime may also be forced by increasing \(N_\varrho\). A higher
density stratification with all the other parameters fixed causes smaller 
length scales and larger velocity amplitudes. This leads to an increase of
\(Ro_\ell\) and to a decrease of \(f_\mathrm{dip}\) at \(Ro_\ell\approx 0.12\)
in Fig. \ref{fig_fdip_nrho}.

\subsection{Equatorial dipole}
\begin{figure}
  \centering
  \includegraphics[scale=0.4]{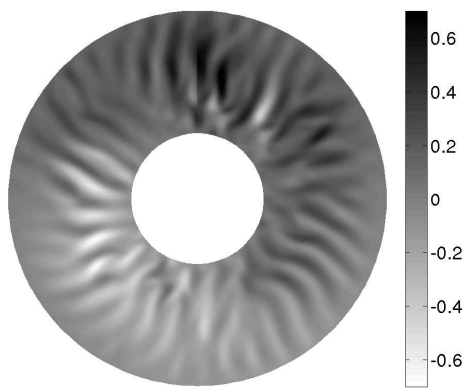}
  \caption{Contour plot (equatorial cut) of the radial magnetic field of 
    model2m at a given time.}
  \label{eq_dipole}
\end{figure}

\begin{figure}
  \centering
  \includegraphics[scale=0.6]{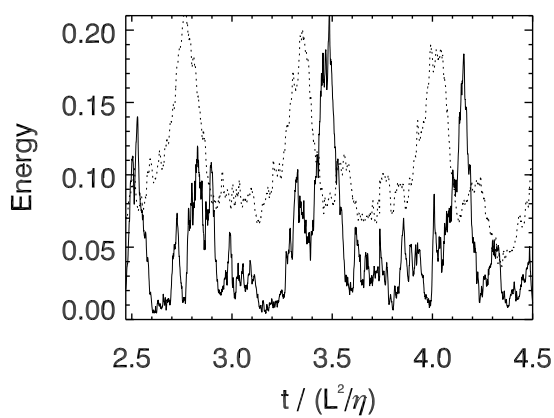}
  \caption{Dotted line: axisymmetric kinetic energy of model5m
  normalised by an arbitrary value; solid line: ratio of axisymmetric 
  to total magnetic energy.}
  \label{fig_energies}
\end{figure}

An example of a model strongly influenced by an equatorial dipole
mode is presented in Fig.~\ref{eq_dipole}. A strong mean zonal flow 
often present in these models seems to be in conflict with the generation of
non-axisymmetric fields.
Figure~\ref{fig_energies} demonstrates that the strong equatorial dipole 
mode of model5m is indeed maintained and rebuilt by the columnar convection 
and damped by the differential rotation. In Fig.~\ref{fig_energies} 
the mean zonal kinetic energy normalised by an arbitrary value (dotted line) 
and the ratio of the axisymmetric magnetic energy to the total magnetic 
energy (solid line) are displayed. The action of the mean zonal flow, or more 
precisely the differential rotation, tends to damp non-axisymmetric components 
of the magnetic field. Thus, a burst of the mean zonal kinetic energy is 
followed by a maximum of the axisymmetric and a dip in the non-axisymmetric 
magnetic energy. Subsequently, the mean zonal flow is quenched by the 
axisymmetric field, the axisymmetric field decays and the non-axisymmetric 
field is rebuilt. The interaction between the mean zonal flow and the 
magnetic field observed in this model is still fairly weak, although the mean 
zonal flow contributes already \(58\%\) to the total kinetic
energy. Therefore, the magnetic field of model5m stays on average 
highly non-axisymmetric. We note that this is very different from the Sun, 
for instance, where probably an even more efficient differential rotation 
causes a predominantly axisymmetric large-scale magnetic field 
\citep{charbonneau10}, but also non-axisymmetric stellar magnetic fields were
reported \citep{donati09}.

\subsection{Discussion}
The fundamental cause of the high dipolarity of dynamo models in the low 
Rossby number regime is an outstanding question. \cite{schrinner12} argued
for Boussinesq models that cylindrical convection in a spherical fluid 
domain leads to a characteristic pattern of the axisymmetric toroidal field 
which eventually results in the clear preference of only one, dipolar eigenmode.
The argument relies on the idea that a line of fluid elements moving towards
the outer spherical boundary has to shorten and causes a converging flow 
towards the equatorial plane. The toroidal field is then advected and 
markedly shaped by this flow component \citep[see also][]{olson99}. This
advection process could be rigorously identified and quantified as a strong
\(\gamma-\)effect in a corresponding mean-field description 
\citep{schrinner07,schrinner12}. In addition, the recent finding that the 
dichotomy of dipolar and multipolar dynamos seems to be absent in 
convective dynamo simulations in Cartesian geometry \citep{tilgner12}
is consistent with this argument and points again to the significance of the 
underlying symmetry constraints. 

What has been said above about Boussinesq models largely applies to 
anelastic models, too. However, geometrical constraints are somewhat relaxed 
for a compressible fluid. Therefore, compressibility might damp the advection 
of the mean toroidal field towards the equatorial plane (\(\gamma-\)effect) 
and we hypothesize that this results in at least two specific differences.

\begin{figure}
  \centering
  \includegraphics[scale=0.6]{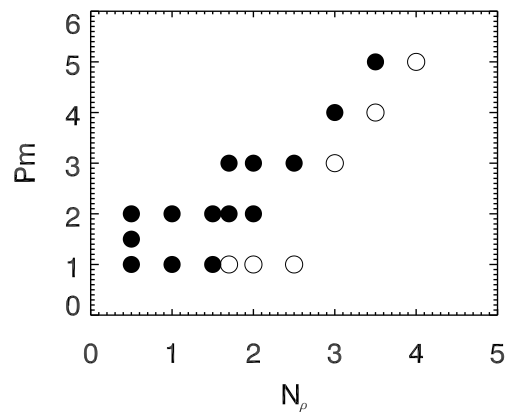}
  \caption{Magnetic Prandtl number versus \(N_\varrho\) for models with 
    \(E=10^{-4}\) and \(Pr=1\) and variable Rayleigh numbers. Filled circles
    stand for parameters for which dipolar solutions were obtained.}
  \label{fig_nrho_pmc}
\end{figure}

First, depending on the {density contrast applied}, it is more difficult to
obtain dipolar solutions for anelastic than for Boussinesq models, even
if \(Ro_\ell<0.12\). However, unlike \cite{gastine12}, we did not 
find that dipolar solutions become impossible if \(N_\varrho\) exceeds a 
certain threshold. {Instead, we observe that for a given}
\(N_\varrho\), Ekman and Prandtl number, there  
seems to {exist} a critical magnetic Prandtl number for dipolar dynamos. 
For \(E=10^{-4}\) and \(Pr=1\), and \(N_\varrho\geq 1.5\), we found 
\(Pm_\mathrm{crit}=2N_\varrho-2\), as apparent from
Fig.~\ref{fig_nrho_pmc}. We emphasize again that the results of
Fig.~\ref{fig_nrho_pmc} depend of course on \(E\) and \(Pr\); the data of our 
numerical study indicate that decreasing \(E\) and increasing \(Pr\) is 
favorable to dipolar dynamo models.%models

Second, magnetic field configurations dominated by an equatorial
dipole seem to be more easily realized in anelastic than in Boussinesq
simulations. For the latter, only a few examples under very specific
conditions were reported \citep{aubert04,gissinger12}. The preference of 
non-axisymmetric modes is well known from dynamo models based on columnar 
convection \citep[e.g.][]{ruediger80,tilgner97}, it is also the case of the
Karlsruhe dynamo experiment \citep{mueller02}.  This agrees
with our reasoning on the importance of the \(\gamma-\)effect in the
axial dipole generation mechanisms \citep[see also][]{schrinner12}. 
Indeed, the \(\gamma-\)effect vanishes in the above examples, as the 
geometrical constraints are relaxed.

\section{Scaling laws}
\label{sec_scaling_laws}
Because of computational limitations, very small length scales
{and time scales associated with extreme} {parameter} {values relevant 
for} planets and stars cannot be resolved in global {direct} numerical 
dynamo simulations. Therefore, numerical models are in 
general not directly comparable to planetary or stellar dynamos. 
Instead, scaling laws, in particular for the field strength, have been derived 
from theory and simulations and then extrapolated to realistic parameter 
regimes \citep[see][and references therein]{christensen10}.

Subsequently, their predictions may 
be compared with planetary or stellar magnetic-field data obtained from 
observations \citep{christensen09,christensen10,davidson13}. 
By this consistency test, scaling laws may provide some evidence about the 
reliability of numerical dynamo models. 

Moreover, different scaling laws typically represent different force 
balances or dynamo mechanisms and their investigation might enable us to better
distinguish between different types of dynamo models. It is in particular this 
second aspect which is of interest in the following. We {adopt here the
approach by \cite{christensen06} and derive scaling laws for the field 
strength, the velocity, the magnetic dissipation time, and the convective heat
transport and compare them} with previous results from Boussinesq 
simulations.  A similar study was recently published by \cite{yadav13} based 
on a somewhat different sample of models. Similarities and differences with 
their findings will be discussed.

{Most of the proposed scaling laws are independent of diffusivities, 
which are thought to be negligible under astrophysical conditions 
\citep{christensen10}}. 
However, present, global dynamo simulations run in parameter regimes 
{where diffusivities still influence} the overall dynamics and weak 
dependencies on the magnetic Prandtl number seem to persist in purely 
empirically derived  scalings 
\citep{christensen04,christensen06,christensen10,yadav12,stelzer13}.  In this
study we do not attempt to resolve this secondary dependence on \(Pm\) 
because the magnetic Prandtl number varies only between 1 and 5 in our sample 
of models.

\subsection{Magnetic field scaling}
\label{sec:mag_field}
The magnetic field strength measured in terms of the Lorentz number scales
with the available energy flux to the power of approximately \(1/3\). We find
for the dipolar dynamos of our sample
\begin{equation}
\frac{Lo}{f_\mathrm{ohm}^{1/2}}\simeq 1.58\,Ra_Q^{0.35} \, ,
\label{scal:2}
\end{equation}
and for the multipolar models
\begin{equation}
\frac{Lo}{f_\mathrm{ohm}^{1/2}}\simeq 1.19\,Ra_Q^{0.34}\, .
\label{scal:4}
\end{equation}
Except for somewhat larger exponential prefactors, this is in good agreement
with previous results from Boussinesq simulations 
\citep{christensen10,schrinner12,yadav12} and very similar to the magnetic
field scaling given by \cite{yadav13}. We note however, that unlike 
\cite{yadav13}, we scale the Lorentz number with the flux based Rayleigh 
number \(Ra_Q\) and not directly with the power released by buoyancy forces. 
Of course, both should be closely related to each other. The same remark 
applies for the velocity scaling discussed below.

Models on the multipolar branch exhibit
lower field strengths compared to their dipolar counterparts. This is not only 
apparent by the smaller prefactor in the multipolar scaling, but 
also the dynamo efficiency \(f_\mathrm{ohm}\) for multipolar models is 
systematically lower than for the corresponding dipolar ones. The latter 
indicates that the bistable behaviour for models at \(Ro_\ell\le 0.12\) 
is caused by different dynamo mechanisms. This was already seen in 
Boussinesq simulations \citep{schrinner12} and later confirmed 
by \cite{gastine12} for anelastic models. 

\begin{figure}
  \centering
  \includegraphics[scale=0.6]{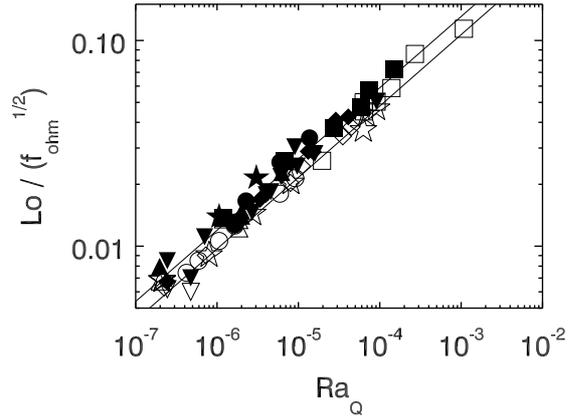}
  \caption{Lorentz number compensated by \(f_\mathrm{ohm}\) versus the 
  flux-based Rayleigh number for our sample of models. Filled symbols
  {correspond to} dipolar models, open symbols are multipolar models and 
  the symbol shape indicates \(N_\rho\) as explained in the caption of 
  Fig. \ref{fig_rol_dip}.}
  \label{fig_ra_lo}
\end{figure}

Apart from a few exceptions, 
the shift between the two scalings in Fig.~\ref{fig_ra_lo} may 
serve to separate dipolar from multipolar dynamos. 
In agreement with \cite{yadav13}, we obtained several models with dipole field 
strengths up to \(f_\mathrm{dip}\approx 0.5\) which nevertheless clearly 
follow the multipolar scaling and belong to the multipolar class of dynamos. 
 
\subsection{Velocity scaling}
\begin{figure}
  \centering
  \includegraphics[scale=0.6]{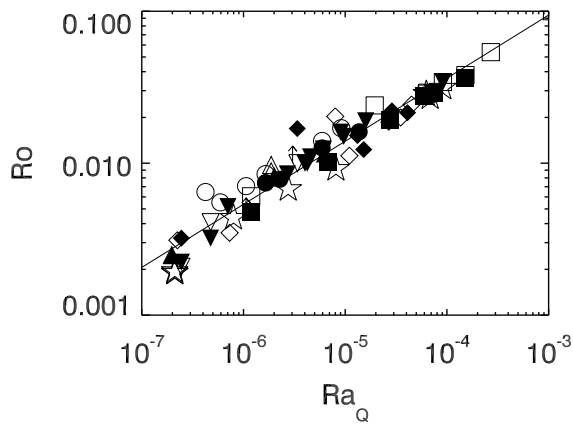}
  \caption{Rossby number versus the flux-based Rayleigh number for our
  sample of models.}
  \label{fig_ra_ro}
\end{figure}

There is an ongoing discussion about the velocity scaling in dynamo models 
\citep{christensen10,davidson13,yadav13}. It is probably not surprising that 
the velocity measured in terms of the Rossby number scales with the flux based
Rayleigh number, but the correct exponent and its theoretical 
justification is debated. The lower bound is set by the assumption of a 
balance between inertia and buoyancy forces (mixing length balance), which
leads to an exponent of \(1/3\) \citep{christensen10}. If however, the 
predominant force balance is assumed to be {between the Lorentz force, 
the buoyancy  and the Coriolis force} (MAC-balance) the exponent is closer 
to \(1/2\) \citep{christensen10,davidson13}. As most previous studies 
\citep{christensen06, christensen10,yadav12,stelzer13,yadav13} we obtained for 
our sample of models an exponent in between {these two values},    
\begin{equation}
Ro=1.66\,Ra_Q^{0.42}\, .
\label{scal:6}
\end{equation}
The scatter in Fig.~\ref{fig_ra_ro} is considerable, but the standard error is 
of the same order as for Boussinesq models with stress-free mechanical boundary 
conditions \citep{yadav12}. Compressible effects do not seem to deteriorate 
the scaling.

However, as in \cite{yadav13}, we are not able to distinguish 
between dipolar and multipolar models in our velocity scaling contrary to 
what has been previously reported by \cite{yadav12} for Boussinesq models. 

\subsection{Scaling of Ohmic dissipation time}
\begin{figure}
  \centering
  \includegraphics[scale=0.6]{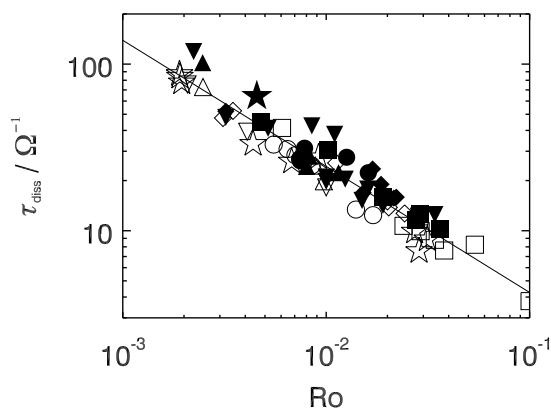}
  \caption{Ohmic dissipation time versus Rossby number for all models
    considered {in this study}.}
  \label{fig_ro_tau}
\end{figure}

The scaling of magnetic dissipation time, 
\begin{equation}
\tau_\mathrm{diss}=E_M/D=\ell_B^2/\eta \, ,
\label{scal:7}
\end{equation}
is used to evaluate the characteristic length scale \( \ell_B\) of the magnetic 
field. \cite{christensen04} originally identified a linear dependence of 
\(\tau_\mathrm{diss}\) on the inverse Rossby number provided that time is 
measured in units of \(\Omega^{-1}\). Their finding was supported by 
dipole-dominated Boussinesq models with no-slip mechanical boundary 
conditions and the evaluation of the Ohmic dissipation time in the Karlsruhe 
dynamo experiment. The best fit for our data points in 
Fig.~\ref{fig_ro_tau}, however, gives an exponent with a significantly lower
absolute value, 
\begin{equation}
\tau_\mathrm{diss}=0.75\,Ro^{-0.76}\, .
\label{scal:8}
\end{equation}
An almost identical result was found by \cite{yadav13} from their 
somewhat more diverse and scattered data set. Apparently, the application of 
stress-free boundary conditions and maybe also compressible effects 
flatten the slope of \(\tau_\mathrm{diss}\) as a function of the Rossby 
number. Moreover, it would seem plausible that \(\tau_\mathrm{diss}\) 
followed different scaling relations for dipolar and multipolar models.
Indeed, for bistable pairs, the dissipation time is systematically larger for 
dipolar than for multipolar models. However, separate least square fits for all 
dipolar and all multipolar models of our sample lead to very similar results.

\subsection{Nusselt number scaling}
\begin{figure}
  \centering
  \includegraphics[scale=0.6]{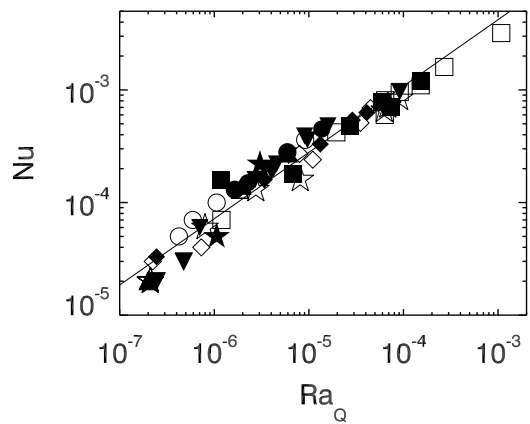}
  \caption{Nusselt number versus the flux-based Rayleigh number for our
  sample of models.}
  \label{fig_ra_nu}
\end{figure}

The convective heat transport in dynamo models is very sensitive to rotation
and depends to a much lower {degree} on the magnetic field, boundary
conditions or the geometry of the fluid domain 
\citep{christensen02,christensen06,aurnou07,schmitz09,busse11,
gastine12a,yadav12,stelzer13}. 
The power law for the Nusselt number inferred from Fig.~\ref{fig_ra_nu},   
\begin{equation}
Nu^\star=0.25\,Ra_Q^{0.59}\, ,
\end{equation}
is consistent with previous {results and confirms this finding
also for anelastic dynamo models;} the exponent of \(0.59\) is very close to 
the value of \(5/9\) established by the above mentioned references. However, 
the scaling is somewhat more scattered than for Boussinesq models 
\citep{yadav12}. We excluded in {a} test all models for which convection is 
{only marginaly above the onset} (\(Nu ^\star<2\)), but this 
{reselection} of models did not  improve the quality of the fit. 

\subsection{Discussion}
\begin{table*}
\caption{Scaling laws for anelastic and Boussinesq models. Results for
  Boussinesq models were taken from \cite{yadav12} 
  \citep[see also][]{schrinner12}.}
\label{table:scaling2}
\centering
\begin{tabular}{lllllll}
\hline \hline
\multicolumn{1}{c}{Scaling}&\multicolumn{3}{c}{Anelastic}&\multicolumn{3}{c}{Boussinesq}\\
 & \multicolumn{1}{c}{c} & \multicolumn{1}{c}{x} & 
\multicolumn{1}{c}{\(\sigma\)} & \multicolumn{1}{c}{c} & \multicolumn{1}{c}{x} 
&\multicolumn{1}{c}{\(\sigma\)}\\
\hline
\(Lo/f_\mathrm{ohm}^{1/2}=c\,Ra_Q^x\)& 1.58&  0.35&0.017 & 1.08&0.37 &0.017\\
    multipolar branch              & 1.19& 0.34& 0.067 & 0.65&0.35 &0.006\\[2mm]
\(Ro=c\,Ra_Q^x\)                   & 1.66&  0.42&0.025 &
0.73\tablefootmark{a}&  0.39\tablefootmark{a}&0.013\tablefootmark{a}\\[2mm]
                                   &     &      &      & 1.79\tablefootmark{b}
& 0.44\tablefootmark{b}&0.010\tablefootmark{b}\\[2mm] 
\(\tau_\mathrm{diss}=c\,Ro^{-x}\)   &  0.75& 0.76 & 0.024
&\multicolumn{1}{c}{--} &0.8&\multicolumn{1}{c}{--}\\[2mm]
\(Nu=c\,Ra_Q^{x}\)                 & 0.25 & 0.59 &0.032 &0.06& 0.52&0.004\\
\hline
\end{tabular}
\tablefoot{\cite{yadav12} distinguished between dipolar and multipolar dynamos
  for their Rossby number scaling, whereas we derived {a single} power law 
  for both classes of dynamo models.
  \tablefoottext{a} {Dipolar models}\tablefoottext{b} {Multipolar models}}  
\end{table*}

In an overall view, the scaling relations for Boussinesq and anelastic models 
are very similar (see Table \ref{table:scaling2}). Beyond that, there is no
obvious effect of compressibility on the scaling results and they might be 
even considered as consistent irrespective of the density stratification of 
the underlying models \citep{yadav13}. However, the reason for the good
agreement could be that the flux-based scaling laws are insensitive to 
different physical conditions. Using the example of the 
magnetic field scaling, we argue in the following that differences in the 
dynamo processes might not be visible in the scaling relation and some caution 
is needed in generalizing results from Boussinesq simulations.  

If the magnetic energy density follows a simple power law in terms of
the convective energy flux an exponent of \(2/3\) is {already required} for 
dimensional reasons \citep[e.g.][]{christensen10}. Moreover, the flux-based 
scaling law for the magnetic field is {composed of the} scalings for the 
velocity and the magnetic dissipation time. By definition,  we have 
\(E_M\sim f_\mathrm{ohm}\,\tau_\mathrm{diss}\,P\) and with 
\(Ro\sim P^\alpha\) and \(\tau_\mathrm{diss}\sim Ro^\beta\), we find 
\(E_M\sim f_\mathrm{ohm}\,P\,P^{\alpha\beta}\). Dimensional arguments require
\(\alpha\,\beta=-1/3\) which establishes relations (\ref{scal:2}) and 
(\ref{scal:4}). Whereas the exponent in the flux-based scaling law for the 
magnetic field is fix, \(\alpha\) and \(\beta\) are to some {extend} 
variable and may change according to the specific physical conditions. This 
reflects {the outcome} of more and more extended parameter studies{:}
the exponent of \(1/3\) in the magnetic-field scaling is reliably reproduced 
but the values for \(\alpha\) and \(\beta\) seem to be less certain and 
are {under} debate.
 
In addition, scaling relations (\ref{scal:2}) and (\ref{scal:4}) require that 
the field strength, measured by \(Lo\), is compensated by the square root
of \(f_\mathrm{ohm}\) \citep[interpreted as dynamo 
efficiency in][]{schrinner13}. However, \(f_\mathrm{ohm}\) probably is a 
complicated function of several control parameters and might depend strongly 
on the specific physical conditions. The often {made} assumption that 
\(f_\mathrm{ohm} \rightarrow 1\) for \(Pm\ll 1\) \citep[e.g.][]{davidson13} 
is probably too simple. For example, \cite{schrinner13} demonstrated recently 
that \(f_\mathrm{ohm}\) in dynamo models might depend strongly on the rotation 
rate. The dynamo efficiency dropped by two orders of magnitude as the rotation 
rate of these models was decreased. A further counterexample could be the 
solar dynamo. Independent estimates result in \(f_\mathrm{ohm}\sim O(10^{-3})\) 
\citep{schrinner13,rempel06} although the magnetic Prandtl number is thought 
to be much smaller than one in the solar interior.\footnote{In
\cite{schrinner13}, a wrong mean solar density has been used to estimate 
\(f_\mathrm{ohm}\) which lead to \(f_\mathrm{ohm}\approx 10^{-2}\). We
correct this error here.} In other words, the flux based scaling laws
probably do not discriminate between different types of dynamos because
differences in the field strength are absorbed by changes in 
\(f_\mathrm{ohm}\).

\section{Conclusions}
\label{sec_conclusions}
Our study revealed a number of similarities between Boussinesq and
anelastic dynamo models. The dichotomy between dipolar and multipolar
models seems to extend to anelastic models, and the flux-based scaling
laws originally proposed for Boussineq models appear to hold similarly
for models in the anelastic approximation. Thus, large scale, dipolar
magnetic fields for both types of models can only be produced if
rotation is important (as measured by the local Rossby number) and the
magnetic field strength is directly related to the energy flux via
(\ref{scal:2}) and (\ref{scal:4}) (see Fig.~\ref{fig_ra_lo}).

However, we also pointed out some significant differences between
Boussinesq and anelastic dynamo simulations. Magnetic field
configurations with a significant equatorial dipole contribution are
less typical for Boussinesq than for anelastic models. Moreover, a
large density stratification in anelastic models may inhibit the
generation of magnetic fields dominated by an axial dipole. The above
claimed consistency of the scalings for Boussinesq and anelastic
simulations partly relies on the very general formulation of the
flux-based scaling laws and does not necessarily imply similar dynamo
processes. We also stress that the assumption of a radially varying
conductivity may introduce additional effects which were not examined
here. Whereas \cite{yadav13} obtained very similar scaling laws for
models with variable conductivities, \cite{duarte13} reported that the
field topology of some models depends on the radial conductivity
profile. A mean-field analysis \citep{schrinner07,schrinner11b} of
numerical dynamo models in the anelastic approximation might give more
detailed insight in relevant dynamo processes and is envisaged for a
future study.

\begin{acknowledgements}
M.S. is grateful for financial support from the DFG fellowship SCHR 1299/1-1.
Computations were performed at CINES, CEMAG, GWDG, and MESOPSL computing 
centres. 
\end{acknowledgements}

\bibliographystyle{aa}
\bibliography{schrinner}
\appendix
\section{Benchmark results}
\label{benchmark}
\begin{table*}[p]
\caption{Hydrodynamic benchmark
  (\(E=10^{-3},\,N_\varrho=5,\,\chi=0.35,\,Ra=3.52\times 10^5,\,Pr=1,\,n=2\))}
\label{table:a1}
\centering
\begin{tabular}{lll}
\hline \hline
Code &PaRoDy&Leeds\\
\hline
K.E. & 81.85 & 81.86\\
Zonal K.E. & 9.388& 9.377\\
Meridional K.E. & 0.02198 & 0.02202\\
&&\\
Luminosity&4.170&4.199\\
&&\\
\(v_\phi\) at \(v_r=0\)&0.8618&0.8618\\
\(S\) at \(u_r=0\)&0.9334&0.9330\\
&&\\
Resolution&\(288\,\times\,192\times\,384\)&\(128\,\times\,192\times\,384\)\\
Timestep&\(5\times10^{-6}\)&\(2.5\times10^{-6}\)\\
\hline
\end{tabular}
\end{table*}

\begin{table*}[p]
\caption{The steady dynamo benchmark (
\(E=2\times 10^{-3},\,N_\varrho=3,\,\chi=0.35,\,Ra=8.00\times 10^4,\,Pr=1,\,Pm=50,\,n=2\))}
\label{table:a2}
\centering
\begin{tabular}{lll}
\hline \hline
Code &PaRoDy&Leeds\\
\hline
K.E. & \(4.189\times 10^6\)& \(4.194\times 10^5\)\\
Zonal K.E. & \(5.993\times 10^4\)& \(6.018\times 10^4\)\\
Meridional K.E. & 52.98& 53.02\\
&&\\
M.E.& \(3.216\times 10^5\)&\(3.202\times 10^5\)\\
Zonal M.E. & \(2.424\times 10^5\)&\(2.412\times 10^5\) \\
Meridional M.E.& \(1.704\times 10^5\)&\(1.697\times 10^4\) \\
&&\\
Luminosity&11.48&11.50\\
&&\\
\(v_\phi\) at \(v_r=0\)&-91.84&-91.78\\
\(B_\theta\) at \(v_r=0\)&\(\pm 0.0343\)&\(\pm 0.03395\)\\
\(S\) at \(u_r=0\)&0.7864&0.7865\\
&&\\
Resolution&\(288\,\times\,126\times\,252\)&\(128\,\times\,144\times\,252\)\\
Timestep&\(5\times10^{-7}\)&\(10^{-6}\)\\
\hline
\end{tabular}
\end{table*}

\begin{table*}[p]
\caption{The unsteady dynamo benchmark (\(E=5\times 10^{-5},\,N_\varrho=3,\,\chi=0.35,\,Ra=2.50\times 10^7,\,Pr=2,\,Pm=2,\,n=2\))}
\label{table:a3}
\centering
\begin{tabular}{lll}
\hline \hline
Code &PaRoDy&Leeds\\
\hline
K.E. & \(2.33\times 10^5\)& \(2.32\times 10^5\)\\
Zonal K.E. & \(1.38\times 10^4\)& \(1.36\times 10^4\)\\
Meridional K.E. & 111& 105\\
&&\\
M.E.& \(2.41\times 10^5\)&\(2.42\times 10^5\)\\
Zonal M.E. & \(9.35\times 10^3\)&\(9.45\times 10^3\) \\
Meridional M.E.& \(2.10\times 10^4\)&\(2.13\times 10^4\) \\
&&\\
Luminosity&42.4&42.5\\
&&\\
Resolution&\(288\,\times\,255\times\,510\)&\(96\,\times\,288\times\,576\)\\
Timestep&\(5\times10^{-7}\)&\(3\times 10^{-6}\)\\
\hline
\end{tabular}
\end{table*}

\section{Critical Rayleigh numbers for the onset of convection}

\begin{table*}[t]
\caption{Overview of the critical Rayleigh numbers and the corresponding 
critical azimuthal wavenumbers for the onset of convection 
determined as explained in \cite{jones_lin}.}
\centering
\begin{tabular}{cccccr}
\hline\hline
\(E\) & \(Pr\) & \(N_\varrho\) & 
\(\chi\) & \(Ra_c\) & \(m_c\) \\
\hline
\(3\times 10^{-5}\)& \(1\)& \(2.0\)& \(0.35\)& \(6.79\times10^6\)&\(24\)\\
\(1\times 10^{-4}\)& \(1\)& \(0.5\)& \(0.35\)& \(3.34\times10^5\)&\(10\)\\
\(1\times 10^{-4}\)& \(1\)& \(1.0\)& \(0.35\)& \(5.67\times10^5\)&\(12\)\\
\(1\times 10^{-4}\)& \(1\)& \(1.5\)& \(0.35\)& \(9.25\times10^5\)&\(14\)\\
\(1\times 10^{-4}\)& \(1\)& \(1.7\)& \(0.35\)& \(1.09\times10^6\)&\(15\)\\
\(1\times 10^{-4}\)& \(1\)& \(2.0\)& \(0.35\)& \(1.43\times10^6\)&\(16\)\\
\(1\times 10^{-4}\)& \(1\)& \(2.5\)& \(0.35\)& \(2.18\times10^6\)&\(19\)\\
\(1\times 10^{-4}\)& \(1\)& \(3.0\)&\(0.35\) & \(3.02\times10^6\)&\(29\)\\
\(1\times 10^{-4}\)& \(1\)& \(3.5\)& \(0.35\)& \(3.62\times10^6\)&\(37\)\\
\(1\times 10^{-4}\)& \(1\)& \(4.0\)& \(0.35\)& \(4.09\times10^6\)&\(43\)\\
\(1\times 10^{-4}\)& \(2\)& \(3.0\)& \(0.35\)& \(7.48\times10^6\)&\(33\)\\
\(1\times 10^{-4}\)& \(1\)& \(2.0\)& \(0.55\)& \(5.35\times10^6\)&\(42\)\\
\(3\times 10^{-4}\)& \(1\)& \(0.5\)& \(0.35\)& \(9.66\times10^4\)&\(7\)\\
\(3\times 10^{-4}\)& \(1\)& \(1.0\)& \(0.35\)& \(1.52\times10^5\)&\(9\)\\
\(3\times 10^{-4}\)& \(1\)& \(1.5\)& \(0.35\)& \(2.32\times10^5\)&\(10\)\\
\(3\times 10^{-4}\)& \(1\)& \(2.0\)& \(0.35\)& \(3.51\times10^5\)&\(12\)\\
\(3\times 10^{-4}\)& \(1\)& \(2.5\)& \(0.35\)& \(5.19\times10^5\)&\(14\)\\
\(3\times 10^{-4}\)& \(1\)& \(3.0\)& \(0.35\)& \(7.12\times10^5\)&\(19\)\\
\(3\times 10^{-4}\)& \(1\)& \(3.5\)& \(0.35\)& \(8.71\times10^5\)&\(25\)\\
\(3\times 10^{-4}\)& \(1\)& \(2.0\)& \(0.45\)& \(6.83\times10^5\)&\(18\)\\
\(3\times 10^{-4}\)& \(1\)& \(2.0\)& \(0.55\)& \(1.26\times10^6\)&\(28\)  \\
\(3\times 10^{-4}\)& \(2\)& \(3.0\)& \(0.35\)& \(8.90\times10^5\)&\(22\)\\
\(1\times 10^{-3}\)& \(1\)& \(2.0\)& \(0.35\)& \(7.70\times10^4\)&\(8\)\\
\(2\times 10^{-3}\)& \(1\)& \(2.5\)& \(0.35\)& \(4.60\times10^4\)&\(7\)\\
\hline
\label{tab:rac}
\end{tabular}
\end{table*}

\section{Numerical models}
\onecolumn
\begin{longtab}[htbp]
\begin{longtable}{ccccccccrcccc}
\caption{Overview of the simulations carried out, ordered with respect to
their local Rossby number.}\\
\hline\hline
Model & \(E\) & \(Ra\) & \(Pm\) & 
\(Pr\) & \(\chi\) & \(N_\varrho\) & 
\(Ro_\ell\) & \(Rm\) & \(Lo\)
& \(f_\mathrm{dip}\) & \(f_\mathrm{ohm}\) & \(Nu\)\\
\hline
\endfirsthead
\caption{continued.}\\
\hline\hline
Model & \(E\) & \(Ra\) & \(Pm\) & 
\(Pr\) & \(\chi\) & \(N_\varrho\) & 
\(Ro_\ell\) & \(Rm\) & \(Lo\)
& \(f_\mathrm{dip}\) & \(f_\mathrm{ohm}\) & \(Nu\)\\
\hline
\endhead
\hline
\endfoot
1m & \(1\times 10^{-4}\)& \(5.00\times 10^6\)& 2.0& 2& 0.35 & 3.0 & 
\(1.54\times 10^{-2}\) & 34 & \(1.10\times 10^{-3}\) & 0.07 & 0.05 & 1.3 \\
2m & \(1\times 10^{-4}\)& \(5.00\times 10^6\)& 3.0& 2& 0.35 & 3.0 &       
\(1.69\times 10^{-2}\) & 57 & \(2.06\times 10^{-3}\) & 0.55 & 0.09 & 1.4 \\
3m & \(1\times 10^{-4}\)& \(5.00\times 10^6\)& 4.0& 2& 0.35 & 3.0 &
\(1.80\times 10^{-2}\) & 78 & \(2.35\times 10^{-3}\) & 0.26 & 0.12 & 1.4 \\
4m & \(1\times 10^{-4}\)& \(5.00\times 10^6\)& 5.0& 2& 0.35 & 3.0 &
\(1.80\times 10^{-2}\) & 95 & \(2.21\times 10^{-3}\) & 0.28 & 0.11 & 1.4 \\
5m & \(1\times 10^{-4}\)& \(2.00\times 10^6\)& 1.0& 1& 0.35 & 0.5 & 
\(2.52\times 10^{-2}\) & 65 & \(2.16\times 10^{-3}\) & 0.43 & 0.08 & 1.5 \\
6d & \(2\times 10^{-3}\)& \(8.00\times 10^4\)& 50 & 1& 0.35 & 3.0 & 
\(2.71\times 10^{-2}\) & 240& \(8.38\times 10^{-3}\)& 0.83 & 0.01 & 1.1 \\
7d & \(5\times 10^{-5}\)& \(1.50\times 10^7\)&2.0 & 1& 0.35 & 1.5 & 
\(2.72\times 10^{-2}\) & 128& \(3.89\times 10^{-3}\)& 0.79 & 0.31 & 1.6 \\
7m & \(5\times 10^{-5}\)& \(1.50\times 10^7\)&2.0 & 1& 0.35 & 1.5 & 
\(2.51\times 10^{-2}\) & 164& \(2.84\times 10^{-3}\)& 0.20 & 0.22 & 1.6 \\
8m & \(1\times 10^{-4}\)& \(2.00\times 10^6\)&1.5 & 1& 0.35 & 0.5 & 
\(2.77\times 10^{-2}\) &  83& \(3.08\times 10^{-3}\)& 0.39 & 0.13 & 1.7 \\
9d& \(3\times 10^{-4}\)& \(4.00\times 10^5\)&3.0 & 1& 0.35 & 0.5 & 
\(2.94\times 10^{-2}\) &  78& \(8.78\times 10^{-3}\)& 0.85 & 0.28 & 1.5 \\
10d& \(1\times 10^{-4}\)& \(7.64\times 10^5\)&2.0 & 1& 0.60 & 1.5 & 
\(3.19\times 10^{-2}\) &  44&\(2.89\times 10^{-3}\) & 0.78 & 0.14 & 1.2 \\
10m& \(1\times 10^{-4}\)& \(7.64\times 10^5\)&2.0 & 1& 0.60 & 1.5 & 
\(3.12\times 10^{-2}\) &  42& \(2.03\times 10^{-3}\)& 0.35 & 0.11 & 1.2 \\
11d& \(1\times 10^{-4}\)& \(2.78\times 10^6\)&2.0 & 1& 0.35 & 1.5 &
\(3.39\times 10^{-2}\) & 104&\(5.72\times 10^{-3}\) & 0.78 & 0.26 & 1.6 \\
12d& \(3\times 10^{-5}\)& \(1.96\times 10^7\)&2.0 & 1& 0.35 & 2.0 &
\(3.39\times 10^{-2}\) & 213&\(3.87\times 10^{-3}\) & 0.68 & 0.32 & 2.1 \\
12m& \(3\times 10^{-5}\)& \(1.96\times 10^7\)&2.0 & 1& 0.35 & 2.0 & 
\(3.33\times 10^{-2}\) & 207&\(3.02\times 10^{-3}\) & 0.24 & 0.23 & 2.0 \\
13d& \(1\times 10^{-4}\)& \(6.14\times 10^6\)&2.0 & 1& 0.60 & 1.0 & 
\(3.41\times 10^{-2}\) &  49&\(2.75\times 10^{-3}\) & 0.68 & 0.12 & 1.2 \\
13m& \(1\times 10^{-4}\)& \(6.14\times 10^6\)&2.0 & 1& 0.60 & 1.0 & 
\(3.44\times 10^{-2}\) &  49&\(2.32\times 10^{-3}\) & 0.05 & 0.12 & 1.2 \\
14m& \(1\times 10^{-4}\)& \(2.50\times 10^6\)&1.5 & 1& 0.35 & 0.5 & 
\(3.51\times 10^{-2}\) & 106&\(4.59\times 10^{-3}\) & 0.49 & 0.19 & 2.0 \\
15d& \(1\times 10^{-4}\)& \(4.01\times 10^6\)&3.0 & 1& 0.35 & 2.5 & 
\(3.76\times 10^{-2}\) & 144&\(6.74\times 10^{-3}\) & 0.52 & 0.24 & 1.7 \\
15m& \(1\times 10^{-4}\)& \(4,01\times 10^6\)&3.0 & 1& 0.35 & 2.5 & 
\(4.51\times 10^{-2}\) & 182&\(6.48\times 10^{-3}\) & 0.47 & 0.23 & 1.7 \\
16d& \(1\times 10^{-4}\)& \(5,00\times 10^6\)&4.0 & 1& 0.35 & 3.0 & 
\(3.76\times 10^{-2}\) & 182&\(6.64\times 10^{-3}\) & 0.47 & 0.23 & 1.5 \\
17d& \(1\times 10^{-4}\)& \(3.00\times 10^6\)&2.0 & 1& 0.35 & 0.5 &
\(3.90\times 10^{-2}\) & 148&\(8.15\times 10^{-3}\) & 0.75 & 0.37 & 2.3 \\
18d& \(1\times 10^{-4}\)& \(3.01\times 10^6\)&1.5 & 1& 0.35 & 0.5 & 
\(4.00\times 10^{-2}\) & 112&\(7.83\times 10^{-3}\) & 0.87 & 0.35 & 2.3 \\
18m& \(1\times 10^{-4}\)& \(3.01\times 10^6\)&1.5 & 1& 0.35 & 0.5 &
\(4.32\times 10^{-2}\) & 127&\(6.09\times 10^{-3}\) & 0.39 & 0.23 & 2.3 \\
19m& \(5\times 10^{-5}\)& \(2.52\times 10^7\)&2.0 & 2& 0.35 & 3.0 &
\(4.46\times 10^{-2}\) & 176&\(4.62\times 10^{-3}\) & 0.30 & 0.26 & 3.4 \\
20d& \(1\times 10^{-4}\)& \(3.40\times 10^6\)&1.0 & 1& 0.35 & 1.0 & 
\(4.71\times 10^{-2}\) &  80&\(8.55\times 10^{-3}\) & 0.86 & 0.37 & 2.4 \\ 
20m& \(1\times 10^{-4}\)& \(3.40\times 10^6\)&1.0 & 1& 0.35 & 1.0 &
\(4.62\times 10^{-2}\) &  97&\(5.93\times 10^{-3}\) & 0.32 & 0.24 & 2.3 \\
21d& \(3\times 10^{-4}\)& \(6.08\times 10^5\)&3.0 & 1& 0.35 & 1.0 & 
\(4.86\times 10^{-2}\) & 115&\(1.27\times 10^{-2}\) & 0.64 & 0.31 & 1.9 \\
22d& \(1\times 10^{-4}\)& \(3.40\times 10^6\) &1.5& 1& 0.35 & 1.0 & 
\(4.88\times 10^{-2}\) & 124&\(8.76\times 10^{-3}\) & 0.75 & 0.35 & 2.4 \\
22m& \(1\times 10^{-4}\)& \(3.40\times 10^6\)&1.5 & 1& 0.35 & 1.0 & 
\(4.78\times 10^{-2}\) & 132&\(6.62\times 10^{-3}\) & 0.39 & 0.25 & 2.3 \\
23m& \(1\times 10^{-4}\)& \(1.00\times 10^7\)&3.0 & 2& 0.35 & 3.0 & 
\(5.13\times 10^{-2}\) & 203&\(6.26\times 10^{-3}\) & 0.14 & 0.19 & 3.6 \\
24d& \(1\times 10^{-4}\)& \(4.00\times 10^6\)& 2.0& 1& 0.35 & 1.5 & 
\(5.30\times 10^{-2}\) & 170&\(8.37\times 10^{-3}\) & 0.69 & 0.33 & 2.6 \\
25d& \(3\times 10^{-4}\)& \(1.00\times 10^6\)& 3.0& 1& 0.35 & 2.5 & 
\(5.36\times 10^{-2}\) & 226&\(1.17\times 10^{-2}\) & 0.75 & 0.21 & 1.6 \\
26m& \(1\times 10^{-4}\)& \(1.14\times 10^7\)&2.0 & 1& 0.60 & 2.0 & 
\(5.38\times 10^{-2}\) &  70&\(3.41\times 10^{-3}\) & 0.24 & 0.14 & 1.4 \\
27m& \(1\times 10^{-4}\)& \(1.00\times 10^7\)&3.0 & 2& 0.35 & 3.0 & 
\(5.46\times 10^{-2}\) & 205&\(6.24\times 10^{-3}\) & 0.21 & 0.19 & 3.6 \\
28d& \(3\times 10^{-4}\)& \(8.00\times 10^5\)&3.0 & 1& 0.35 & 0.5 & 
\(6.06\times 10^{-2}\) & 161&\(2.14\times 10^{-2}\) & 0.76 & 0.41 & 2.5 \\
29d& \(1\times 10^{-4}\)& \(5.00\times 10^6\)&3.0 & 1& 0.35 & 2.0 & 
\(6.10\times 10^{-2}\) & 270&\(1.00\times 10^{-2}\) & 0.61 & 0.36 & 2.6 \\
30m& \(1\times 10^{-4}\)& \(5.00\times 10^6\) &1.0 & 1& 0.35 & 1.7 & 
\(6.50\times 10^{-2}\) & 100&\(8.06\times 10^{-3}\) & 0.5  & 0.26 & 2.6 \\
31d& \(1\times 10^{-4}\)& \(5.00\times 10^6\) &1.0 & 1& 0.35 & 0.5 &
\(6.70\times 10^{-2}\) & 126&\(1.41\times 10^{-2}\) & 0.83 & 0.48 & 3.8 \\
31m& \(1\times 10^{-4}\)& \(5.00\times 10^6\) &1.0 & 1& 0.35 & 0.5 & 
\(7.10\times 10^{-2}\) & 140&\(1.05\times 10^{-2}\) & 0.45 & 0.34 & 3.8 \\
32d& \(1\times 10^{-4}\)& \(5.00\times 10^6\) &2.0 & 1& 0.35 & 1.5 & 
\(6.70\times 10^{-2}\) & 220&\(1.16\times 10^{-2}\) & 0.65 & 0.40 & 3.2 \\
33d& \(1\times 10^{-4}\)& \(5.00\times 10^6\) &3.0 & 1& 0.35 & 1.7 & 
\(6.70\times 10^{-2}\) & 300&\(1.08\times 10^{-2}\) & 0.63 & 0.37 & 2.9 \\
34d& \(1\times 10^{-4}\)& \(5.00\times 10^6\) &2.0 & 1& 0.35 & 1.7 & 
\(6.80\times 10^{-2}\) & 200&\(1.07\times 10^{-2}\) & 0.67 & 0.35 & 3.0 \\
35d& \(3\times 10^{-4}\)& \(6.08\times 10^5\) &3.0 & 1& 0.35 & 1.5 & 
\(6.99\times 10^{-2}\) & 160&\(1.78\times 10^{-2}\) & 0.59 & 0.34 & 2.3 \\
36d& \(1\times 10^{-4}\)& \(8.00\times 10^6\) &4.0 & 1& 0.35 & 3.0 & 
\(7.00\times 10^{-2}\) & 380&\(1.41\times 10^{-2}\) & 0.62 & 0.40 & 2.9 \\
37d& \(3\times 10^{-4}\)& \(1.05\times 10^6\) &3.0 & 1& 0.35 & 2.0 & 
\(7.21\times 10^{-2}\) & 152&\(1.60\times 10^{-2}\) & 0.67 & 0.31 & 2.1 \\
38m& \(1\times 10^{-4}\)& \(1.20\times 10^7\) &5.0 & 1& 0.35 & 4.0 & 
\(7.22\times 10^{-2}\) & 458&\(1.15\times 10^{-2}\) & 0.19 & 0.32 & 2.6 \\
39d& \(1\times 10^{-4}\)& \(5.55\times 10^6\) &2.0 & 1& 0.35 & 1.5 & 
\(7.40\times 10^{-2}\) & 247&\(1.52\times 10^{-2}\) & 0.63 & 0.49 & 3.6 \\
40d& \(1\times 10^{-4}\)& \(6.00\times 10^6\) &1.0 & 1& 0.35 & 0.5 & 
\(8.00\times 10^{-2}\) & 150&\(1.64\times 10^{-2}\) & 0.82 & 0.49 & 4.7 \\
41d& \(1\times 10^{-4}\)& \(1.00\times 10^7\) &5.0 & 1& 0.35 & 3.5 & 
\(8.20\times 10^{-2}\) & 500&\(1.23\times 10^{-2}\) & 0.63 & 0.33 & 3.2 \\
42d& \(1\times 10^{-4}\)& \(6.50\times 10^6\) &1.0 & 1& 0.35 & 1.5 & 
\(8.80\times 10^{-2}\) & 150&\(1.64\times 10^{-2}\) & 0.77 & 0.45 & 4.6 \\
43d& \(1\times 10^{-4}\)& \(7.00\times 10^6\) &2.0 & 1& 0.35 & 2.0& 
\(8.80\times 10^{-2}\) & 280&\(1.64\times 10^{-2}\) & 0.66 & 0.46 & 4.3 \\
44m& \(1\times 10^{-4}\)& \(7.00\times 10^6\) &1.0 & 1& 0.35 & 2.0 & 
\(9.05\times 10^{-2}\) & 140&\(1.18\times 10^{-2}\) & 0.04 & 0.34 & 3.7 \\
45d& \(1\times 10^{-3}\)& \(2.50\times 10^5\) &3.0 & 1& 0.35 & 2.5 & 
\(9.07\times 10^{-2}\) &  87&\(2.76\times 10^{-2}\) & 0.82 & 0.23 & 1.7\\
45m& \(1\times 10^{-3}\)& \(2.50\times 10^5\) &3.0 & 1& 0.35 & 2.5 & 
\(6.35\times 10^{-2}\) &  86&\(2.17\times 10^{-2}\) & 0.31 & 0.19 & 1.6 \\
46m& \(1\times 10^{-4}\)& \(6.00\times 10^6\) &1.0 & 1& 0.35 & 0.5 & 
\(9.20\times 10^{-2}\) & 170&\(1.26\times 10^{-2}\) & 0.52 & 0.35 & 4.6 \\
47d& \(3\times 10^{-4}\)& \(1.50\times 10^6\) &3.0 & 1& 0.35 & 2.5 & 
\(9.35\times 10^{-2}\) & 191&\(2.29\times 10^{-2}\) & 0.59 & 0.37 & 2.6 \\
47m& \(3\times 10^{-4}\)& \(1.50\times 10^6\) &3.0 & 1& 0.35 & 2.5 & 
\(8.67\times 10^{-2}\) & 234&\(1.67\times 10^{-2}\) & 0.39 & 0.22 & 2.2 \\
48m& \(3\times 10^{-4}\)& \(2.52\times 10^6\) &3.0 & 1& 0.55 & 2.0 & 
\(9.81\times 10^{-2}\) & 112&\(1.12\times 10^{-2}\) & 0.27 & 0.21 & 1.8 \\
49d& \(3\times 10^{-4}\)& \(1.40\times 10^6\) &3.0 & 1& 0.35 & 2.0 & 
\(1.01\times 10^{-1}\) & 221& \(2.47\times 10^{-2}\)& 0.51 & 0.36 & 2.8 \\
50d& \(3\times 10^{-4}\)& \(1.50\times 10^6\) &3.0 & 1& 0.35 & 1.5 &
\(1.09\times 10^{-1}\) & 265&\(3.18\times 10^{-2}\) & 0.67 & 0.45 & 3.7 \\
51d& \(1\times 10^{-4}\)& \(8.00\times 10^6\) &1.0 & 1& 0.35 & 1.5 & 
\(1.10\times 10^{-1}\) & 190&\(1.97\times 10^{-2}\) & 0.77 & 0.49 & 5.8 \\
52d& \(1\times 10^{-4}\)& \(1.52\times 10^7\) &3.0 & 1& 0.55 & 2.0 & 
\(1.13\times 10^{-1}\) & 369&\(2.10\times 10^{-2}\) & 0.64 & 0.54 & 4.3 \\
53d& \(3\times 10^{-4}\)& \(2.05\times 10^6\) &3.0 & 1& 0.45 & 2.0 & 
\(1.14\times 10^{-1}\) & 186&\(2.27\times 10^{-2}\) & 0.60 & 0.35 & 2.6 \\
54d& \(1\times 10^{-3}\)& \(3.00\times 10^5\) &3.0 & 1& 0.35 & 2.5 & 
\(1.19\times 10^{-1}\) & 108&\(4.14\times 10^{-2}\) & 0.74 & 0.33 & 2.2 \\
54m& \(1\times 10^{-3}\)& \(3.00\times 10^5\) &3.0 & 1& 0.35 & 2.5 & 
\(1.13\times 10^{-1}\) & 114&\(2.97\times 10^{-2}\) & 0.27 & 0.28 & 2.1 \\
55d& \(3\times 10^{-4}\)& \(2.00\times 10^6\) &3.0 & 1& 0.35 & 2.5 & 
\(1.22\times 10^{-1}\) & 277&\(2.91\times 10^{-2}\) & 0.56 & 0.38 & 3.6 \\
56d& \(3\times 10^{-4}\)& \(2.39\times 10^6\) &3.0 & 1& 0.45 & 2.0 & 
\(1.29\times 10^{-1}\) & 215&\(2.73\times 10^{-2}\) & 0.59 & 0.42 & 3.1 \\
56m& \(3\times 10^{-4}\)& \(2.39\times 10^6\) &3.0 & 1& 0.45 & 2.0 & 
\(1.25\times 10^{-1}\) & 243&\(2.05\times 10^{-2}\) & 0.32 & 0.29 & 3.3 \\
57m& \(3\times 10^{-4}\)& \(2.08\times 10^6\) &3.0 & 1& 0.35 & 2.5 & 
\(1.30\times 10^{-1}\) & 291&\(2.71\times 10^{-2}\) & 0.32 & 0.36 & 3.7 \\
58m& \(3\times 10^{-4}\)& \(2.50\times 10^6\) &3.0 & 1& 0.35 & 3.0 & 
\(1.28\times 10^{-1}\) & 272&\(2.58\times 10^{-2}\) & 0.26 & 0.35 & 3.4 \\
59m& \(3\times 10^{-4}\)& \(5.00\times 10^5\) &2.0 & 2& 0.35 & 3.0 & 
\(1.34\times 10^{-1}\) & 191&\(1.98\times 10^{-2}\) & 0.28 & 0.29 & 5.4 \\
60m & \(3\times 10^{-4}\)& \(3.50\times 10^6\) &3.0 & 1& 0.35 & 3.5 & 
\(1.41\times 10^{-1}\) & 294&\(2.58\times 10^{-2}\) & 0.17 & 0.34 & 3.6 \\
61m& \(3\times 10^{-4}\)& \(2.85\times 10^6\) &3.0 & 1& 0.35 & 3.0 & 
\(1.43\times 10^{-1}\) & 314&\(2.79\times 10^{-2}\) & 0.21 & 0.36 & 3.8 \\
62m& \(3\times 10^{-4}\)& \(2.50\times 10^6\) &3.0 & 1& 0.35 & 2.5 & 
\(1.47\times 10^{-1}\) & 342&\(3.15\times 10^{-2}\) & 0.28 & 0.39 & 4.2 \\
63m& \(1\times 10^{-4}\)& \(1.10\times 10^7\) &1.0 & 1& 0.35 & 2.5 & 
\(1.50\times 10^{-1}\) & 240&\(1.58\times 10^{-2}\) & 0.28 & 0.37 & 5.2 \\
64m& \(3\times 10^{-4}\)& \(3.78\times 10^6\) &3.0 & 1& 0.55 & 2.0 & 
\(1.52\times 10^{-1}\) & 200&\(1.92\times 10^{-2}\) & 0.17 & 0.31 & 2.7 \\
65m& \(1\times 10^{-3}\)& \(4.00\times 10^5\) &3.0 & 1& 0.35 & 2.5 & 
\(1.66\times 10^{-1}\) & 162&\(4.39\times 10^{-2}\) & 0.26 & 0.26 & 2.6 \\
66m& \(2\times 10^{-3}\)& \(2.00\times 10^5\) &3.0 & 1& 0.35 & 2.5 &
\(2.18\times 10^{-1}\) & 150&\(4.66\times 10^{-2}\) & 0.19 & 0.17 & 2.6 \\
\label{tab:2}
\end{longtable}
\tablefoot{In case of bistable pairs, the local
  Rossby number of the dipolar model was considered to create the
  sequence. Kinematic stability: out of the 13 models tested, kinematically
  stable are model16d, model37d, model45d, model54d, model55d, and model56d,
  and kinematically unstable are model2m, model5m, model8m, model14m,
  model18m, model65m, and model66m. Models with a considerable equatorial
  dipole field are: model2m, model5m, model8m, model14m, model18m, model22m, 
  model27m, model30m, model31m, model45m, model46m, and model56m.}    
\end{longtab}

\end{document}